\RequirePackage{fix-cm}
\documentclass[twocolumn]{svjour3}          
\smartqed  
\usepackage{graphicx}
\usepackage{enumerate}
\usepackage{graphics}
\usepackage{subfigure}
\usepackage{stfloats}
\usepackage{caption}
\usepackage{array}
\usepackage{url}
\usepackage{extarrows}
\usepackage{subfloat}

\begin{document}

\title{NetSecCC: A Scalable and Fault-tolerant Architecture without Outsourcing Cloud Network Security
}


\author{Jin~He \and Mianxiong~Dong \and  Kaoru~Ota \and Minyu~Fan \and Guangwei~Wang}


\institute{J. He, M. Fan and G. Wang \at
Department of Computer Science, University of Electronic Science and Technology of China (UESTC), Chengdu, 611731, P.R.China\\
              \email{\{hejin\_some, ff98, wguanwei\}@163.com}           
           \and
           M. Dong \at
           National Institute of Information and Communications Technology, Japan\\
            \email{mx.dong@nict.go.jp}
           \and
           K. Ota \at
           Department of Information and Electronic Engineering, Muroran Institute of Technology, Japan\\
            \email{ota@csse.muroran-it.ac.jp}
}

\date{Received: date / Accepted: date}

\maketitle

\begin{abstract}
Modern cloud computing platforms based on virtual machine monitors carry a variety of complex business that present many network security vulnerabilities. At present, the traditional architecture employs a number of security devices at front-end of cloud computing to protect its network security. Under the new environment, however, this approach can not meet the needs of cloud security. New cloud security vendors and academia also made great efforts to solve network security of cloud computing, unfortunately, they also cannot provide a perfect and effective method to solve this problem. We introduce a novel network security architecture for cloud computing (NetSecCC) that addresses this problem. NetSecCC not only provides an effective solution for network security issues of cloud computing, but also greatly improves in scalability, fault-tolerant, resource utilization, etc. We have implemented a proof-of-concept prototype about NetSecCC and proved by experiments that NetSecCC is an effective architecture with minimal performance overhead that can be applied to the extensive practical promotion in cloud computing.

\end{abstract}

\section{Introduction}
\label{intro}
computing is a new computing paradigm that
is built on distributed and parallel computing, virtualization, network storage technologies, load balance,
utility computing, and service-oriented architecture.
In the last several years, cloud computing has emerged as one
of the most influential paradigms in the IT industry, and has
attracted extensive attention from both academia and industry. The
benefits of cloud computing include reduced costs and capital
expenditures, increased operational efficiencies, scalability,
flexibility, immediate time to market, and so on.

Although the great benefits brought by cloud computing paradigm
are exciting for IT companies, academic researchers, and
potential cloud users, cloud security becomes serious obstacles which, without being appropriately addressed, will prevent cloud computing's extensive applications and usage in the future. Especially, cloud network security has become one of the prominent security concerns \cite{fernandes2013security} \cite{networksecurity-cloud-1} \cite{zissis2012addressing} \cite{heiser2008assessing} \cite{popovic2010cloud} \cite{mell2011nist} \cite{wu2013towards} \cite{zhu2014probabilistic} \cite{pearce2013virtualization}, even the vast majority of data destruction or tampering or forgery in cloud computing mainly come from malicious network attacks \cite{mohammed2012enhancing}. It is further evidence from National Vulnerability Database (NVD) \cite{NVD} that until February 2013, 84 network vulnerabilities have been discovered in cloud computing. All the above evidence has strongly confirmed that malicious attacks from network is a serious security threat to a variety of network-based services (e.g., Website, bank, date center) in cloud computing.

There are multiple ways to solve network security issues of cloud computing, we divided them into three types: the solution from the traditional architecture, the solution from cloud provider and recent efforts from academia research. The traditional architecture as shown in Fig 1 places network security devices (middleboxs \cite{joseph2008modeling}) at front-end of cloud computing to protect their network security, but the architecture is applied to cloud security to bring some problems. \textbf{\emph{Is lack of network security protection between VMs}}: Since a compromised VM easily attacks other VMs in the same hardware platform by virtual network \cite{duncan2012insider} \cite{HyperLock}, cloud security is required to prevent not only malicious attacks from external traffic but also internal attacks from the malicious VM, thereby ensuring network security of cloud computing. However, the traditional architecture is lack of internal network protection mechanism between VMs. \textbf{\emph{Difficult scalability}}: The traditional architecture appears such a scenario: traffic bursts and exceeds the maximum capacity of the existing deployed middleboxs at some point, while traffic in other times is in the normal work. If we add the corresponding middleboxs to avoid traffic loss by peak load, thus resulting in not only less efficient resource utilization, but also higher costs and more post-maintenance costs. \textbf{\emph{Difficult Fault-tolerance:}} Using  hot standby (HS) in the traditional architecture can offer fault tolerance for the failed middelboxs. However, only a enterprise network requires 640 middleboxs to protect its security \cite{sekar2012design} \cite{sekar2011middlebox}, not to mention, cloud computing hosting more complex multi-services needs much more middleboxs than a enterprise network. If we also use the same hot standby to offer fault tolerance in cloud computing for such a large-scale middleboxs, this results in unsustainable costs.

\begin{figure}[!ht]
\centering
\includegraphics[width=2.8in]{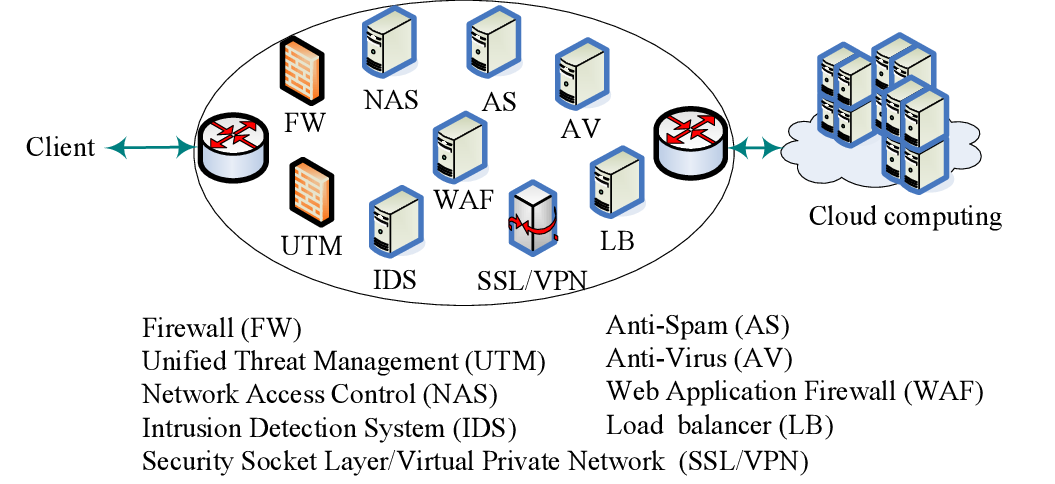}
\caption{The traditional architecture}
\label{fig_sim1}
\end{figure}

In order to address the shortcomings of the traditional architecture and provide suitable for network security service of cloud computing, cloud security vendors have taken some measures. McAfee SaaS \cite{McAfee-saas} merely provides for a single type of security protection in cloud computing (Web and Email), but is lack of complex multi-service protection in cloud. Amazon Web Services (AWS) \cite{AWS} only provide basic network security with a port-based firewall, they need to turn to partners like security vendors to provide robust network security with the granularity, control and reporting that you need. VMware vShield (app, endpoint, edge, zones) \cite{vmware} provides services in cloud with partial network security protection, but is short of comprehensive and integrated capacity (e.g., encryption transmission, anti-virus). Security as a Service (SecaaS) \cite{SecaaS} provides services in cloud with comprehensive security protection, including web, email and intrusion SecaaS, but does not involve their scalability, system fault-tolerance and cooperation between them.

In academia, Wu \emph{et al.} \cite{wu2010network} aim to control the inter-communication among virtual machines with higher security by the embedded firewall in virtualized environment, but this method does not prevent malicious attacks from external traffic, and also not involves in flexible scalability and fault tolerance for the firewall. Salah \emph{et al.} \cite{salah2013using} have proposed cloud-based security overlay network which can provide a comprehensive protection solution for servers and end-users, but is also lack of an effective scalability and fault-tolerance mechanism. Split/Merge \cite{rajagopalan2013split} can be dynamically scaled out (or in) virtual middleboxs in cloud computing by SDN \cite{openflow}, which only focuses on load-balanced elasticity and system utilization without paying attention to preventing external and internal malicious traffic from attacking on cloud services. \cite{sekar2011middlebox} \cite{qazi2013simple} well combine with middleboxs and SDN to protect enterprise network security, and provides a flexible scalability and fault-tolerance mechanism, but it's a pity that they are not suitable for cloud security.

Since it is not suitable or defective for the above efforts to protect network security of cloud computing, we propose NetSecCC architecture that takes a novel approach of eliminating these disadvantages. It not only prevents external and internal malicious attacks and offer on-demand network security service for cloud users, but also is able to provide flexible scalability and fault tolerance for virtual middleboxs' load and failure, respectively. Experiments have further fully proved that NetSecCC has efficient results in terms of scalability and fault tolerance, and also provides security services for cloud computing without sacrificing great performance as consideration.
In summary, our main contributions are as follows:
\begin{itemize}
  \item \textbf{\emph{Innovative architecture}} We propose a novel flexible effective security architecture which uses a systematic approach to properly provide security protection for cloud computing, and to guide cloud computing road to industrialization.
  \item \textbf{\emph{Preventing external and internal malicious attacks }} NetSecCC not only protects against malicious attacks from external traffic, but also prevents attacks from internal traffic to ensure network security of cloud users' services in cloud computing.
  \item \textbf{\emph{Scalability}} Our architecture presents balanced scalability alongside VM scale-in and scale-out for virtual middleboxs according to their load.
  \item \textbf{\emph{Fault tolerance}} When VMs hosting virtual middleboxs fail, our approach provides many-to-one fault-tolerant mechanism to overcome disadvantages of the traditional HS.
\end{itemize}
\begin{figure*}[!ht]
\centering
\includegraphics[width=5.0in]{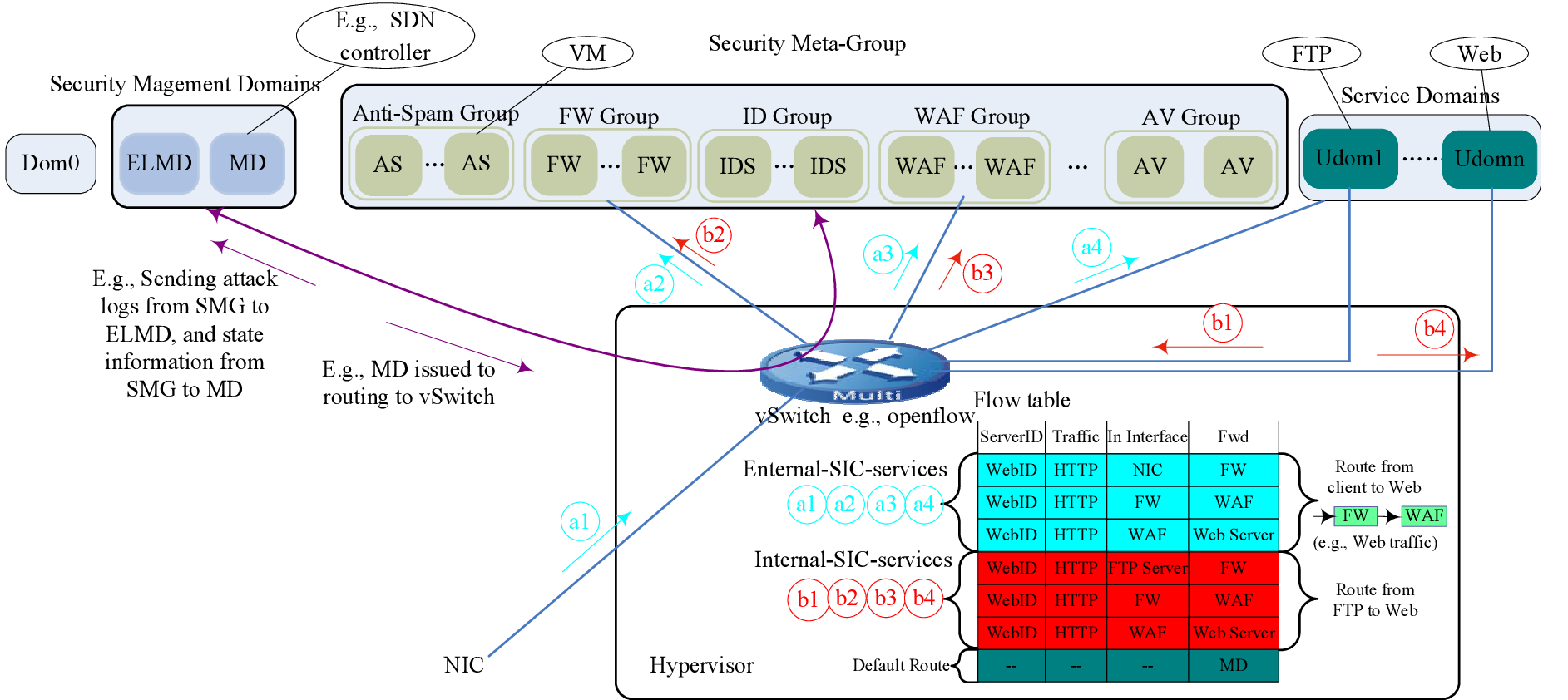}
\caption{NetSecCC architecture}
\label{fig_sim1}
\end{figure*}
The rest of the paper is organized as follows: Section 2 provides an overview of the design of NetSecCC. Section 3 details the implementation of the entire system. In section 4, we show the results of the experiments we conducted for evaluating
the impact and performance of our system. Finally, Section 5 is the conclusion.

\section{Design}
\label{sec:1}
Before we describe the NetSecCC design, we assume that hardware platform, hypervisor and VMs-OS on cloud computing are trusted, we only focus on network security of services placed on cloud computing. Fig. 2 shows NetSecCC architecture, security manage domain (SMD) as a controller guides external and internal traffic through security Meta-Group (SMG) to be filtered and inspected by vSwitch, SMG as a performer performs security inspection and filtering for incoming and outgoing traffic of cloud users' services, service domains host cloud users' services. Next, we present the components of NetSecCC and their operations.

\subsection{Principle}
As Fig 2 shows, NetSecCC mainly consists of four parts: a system domain (dom0), security management domains, security meta-group and service domains. We first begin by the components of NetSecCC and their operations before elaborating on NetSecCC principle.

\begin{itemize}
\item \emph{\texttt{Dom0}} We weaken dom0 privileges, it does not have permission to create/start and stop/destroy any domain in SMG. At the same time, dom0 still keeps such permissions to do with all domains in service domains and SMD, and manages resources, including scheduling time-slices, I/O quotas, etc.
\item \emph{\texttt{SMD}} is composed of management domain (MD) and event and log management domain (ELMD). MD is responsible for three main functions: first, create/destroy any domain in SMG; Second, collect state information (e.g., CPU utilization, sessions) from every group in SMG and receive state information (e.g., load, failure) from vSwich; Third, generate and update routing in virtual switch (vSwtich) according to security inspection chains (SIC) \S(2.2), virtual middleboxs' load and failure. ELMD stores and manages events and logs from SMG, and provides the unified query for security managers.
\item \emph{\texttt{SMG}} is comprised of various security meta-groups (e.g., WAF group, IDS group, AV group, etc). Every group includes one or multiple virtual middleboxs (To simplify, we also call virtual middleboxs security domains) such as IDS, AV. Note that each virtual middlebox is installed in a standalone VM. Security domains are responsible for traffic security inspection and filtering, and provide fault-tolerant for the failed security domains by the improved Hot Standby (HS).
\item \emph{\texttt{Service Domains}} host various types of Internet-based cloud users's services (e.g., FTP server, Web server).
\item \emph{\texttt{vSwtich}} is responsible for receiving routing from MD, and forwarding external and internal traffic through security domains to be filtered and inspected.
\end{itemize}

Next, in order to clearly describe NetSecCC work principle, we divides it into three steps:
\begin{itemize}
\item \emph{\texttt{Generate routing}} MD as a SDN controller generates and issues to routing to vSwitch according to SIC mapped by network security requirements of cloud users' services (To simplify, we call it SIC of cloud users' services), security domains topology, middleboxs' load and failure from SMG. Specifically, MD generates and updates routing in vSwtich in the following two stages: The initial phase, when cloud users employ their services in service domains before not running, MD generates routing in accordance with SIC of cloud users' service, security domains topology and current middleboxs' load; The running phase, when middelboxs appears overload and low-load or failure, MD updates routing in vSwtich to rebalance middleboxs's load for overload and low-load, and provide fault tolerance for failure.
\item \emph{\texttt{Forward traffic}} VSwitch as a openflow switcher forwards external and internal traffic through SMG to be inspected and filtered according to routing in vSwitch. Incompetence external traffic from Internet or internal traffic from VMs must route through SMG before arriving at services in service domains, thereby ensuring services security. To make this process concrete, we use Web server in service domains as an example as shown in Fig 2 to elaborate on the processing. SIC (FW-WAF) of Web server can guarantee that internal and external traffic arriving at Web server is secure and trusted. Web traffic is first forwarded to FW by vSwtich to be filtered, then forwarded to WAF to be inspected, finally, arrives at Web server. Here vSwitch contains forwarding routing mapped by Web security chains FW-WAF, light blue areas in flow table express routing from external traffic through chains FW-WAF, red areas indicate routing from internal traffic through FW-WAF.
\item \emph{\texttt{Filtering and inspection}} SMG is responsible for filtering and inspecting incoming traffic before it is forwarded to service domains. According to the corresponding SIC, traffic is required to go through one or more security groups to ensure that traffic arriving at service domains is secure and trusted. That is, first group on SIC path receives and performs traffic security inspection, then forwards it to vSwtich. If SIC has next group, traffic is forwarded to it to be filtered and inspected, in turn until the last group.
\end{itemize}

We can observe from NetSecCC work principle that the corresponding SIC of cloud users' services and scalability and fault-tolerance of every group in SMG are the focus of NetSecCC design. SIC of cloud users' services focuses on on-demand security service (\S 2.2), while every group shows flexible scalability and efficient fault tolerance (\S 2.3) to increase load balancing and high availability, and improve resource utilization.

\subsection{SIC}
\label{sec:2}
\begin{figure}[!ht]
\centering
\includegraphics[width=2.6in]{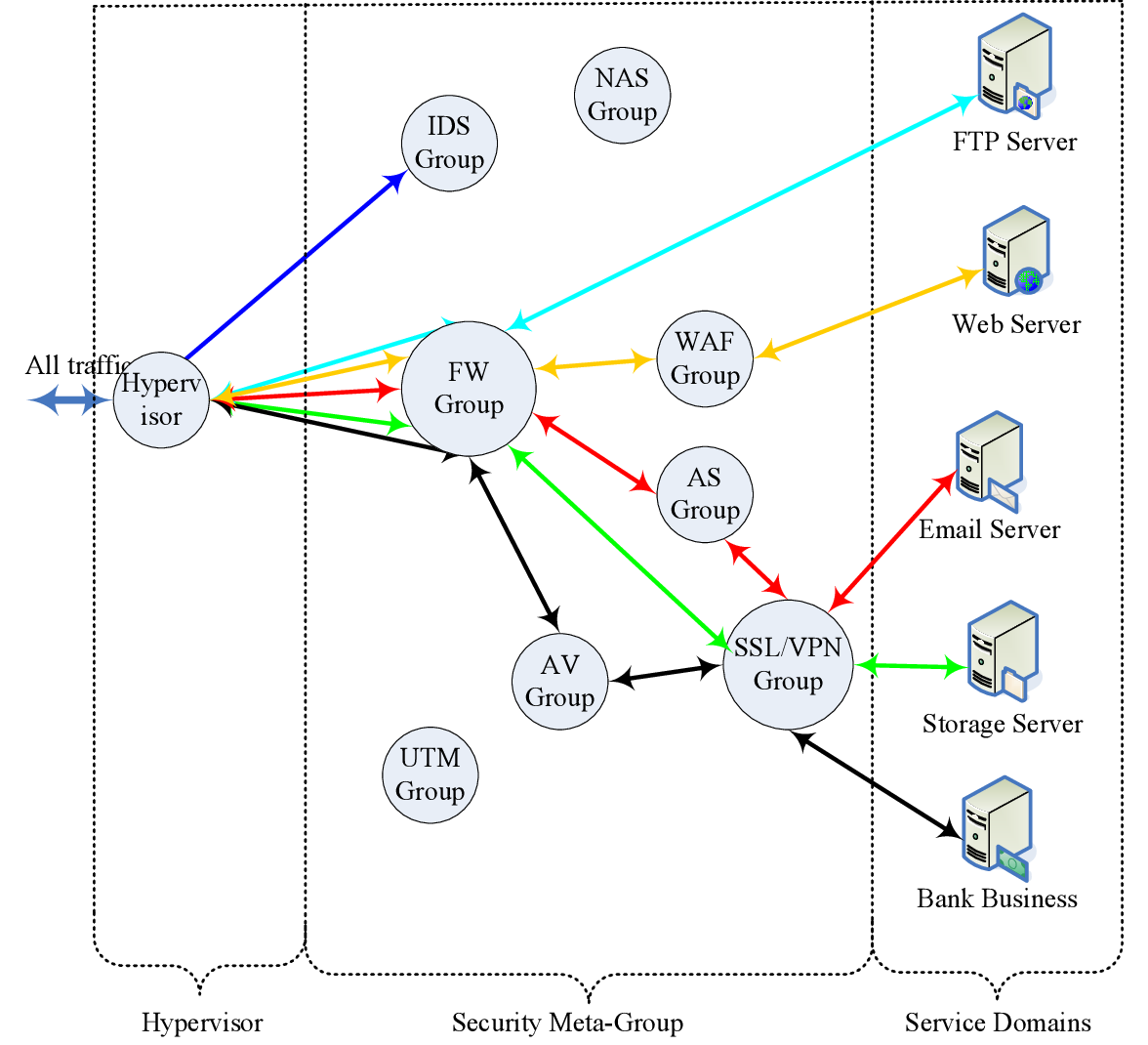}
\caption{Security Inspection Chains}
\label{fig_sim4}
\end{figure}
SIC is a sequence of logical policy chains through one or more security groups (e.g., FW-WAF, FW-IDS), traffic accessing to service domains must route through the corresponding SIC to ensure the security  of service domains. NetSecCC is able to offer suitable for their SICs according to different network security requirements from different services, namely, on-demand security service as shown in Fig 3. Note that many middleboxes are stateful and need to process both directions of a session for correctness. To make this discussion concrete, we use two examples to further illustrate SIC.

\textbf{\emph{Web server}} in service domains needs to solve these attacks from network-layer and application-layer. Attacks from network-layer include DDOS attack, syn attack, etc. Attacks from application-layer includes cross-site attacks, SQL injection, vulnerability overflow and so on. NetSecCC provides Web server security with SIC (FW-WAF) as shown in Fig 3 using yellow lines, Web traffic must flow through FW and WAF, where FW group assures its network-lay security, WAF group offers its application-layer security, thereby ensuring that traffic reaching web service is secure.

\textbf{\emph{Email server}} security requirements are able to prevent from DDOS attack, syn attack, malicious e-mail, spam and virus e-mail, etc. Even some important emails need to be encrypted transmission. NetSecCC provides email server with SICs (FW-AS-SSL/VPN) indicated in Fig 3 with red lines to guarantee its security. Where FW group secures network-layer security of email server, AS group filter malicious and spam e-mail to guarantee application-layer security, and SSL/VPN group provides some important emails with secure transmission.

\subsection{Group Management}
\label{sec:3}
MD as a SDN controller is responsible for controlling traffic accessing to services in service domains to follow their corresponding SICs. While each group on SIC path is a real performer on security inspection and filtering, preventing malicious and virus attacks to ensure that external and internal traffic arriving at services traffic are secure and trusted. In this process, when security domains (nodes) in some group on SIC path appear overload and low-load and failure, NetSecCC needs to rebalance load in groups for overload and low-load to strengthen network traffic processing capability, including increasing throughput and resource utilization, and to provide fault tolerance using hot standby for failure to improve seamless inspection and filtering, including reducing system recover time. Note that the HS is not a traditional one-to-one relationship \cite{HA-f5} \cite{HA-Vyatta} \cite{zhu2013zoom} between actives node and standby nodes, but is a improved many-to-one relationship, namely, state information of all active nodes are synchronized to one standby to improve resource utilization.

\begin{figure}[!ht]
\centering
\includegraphics[width=2.5in]{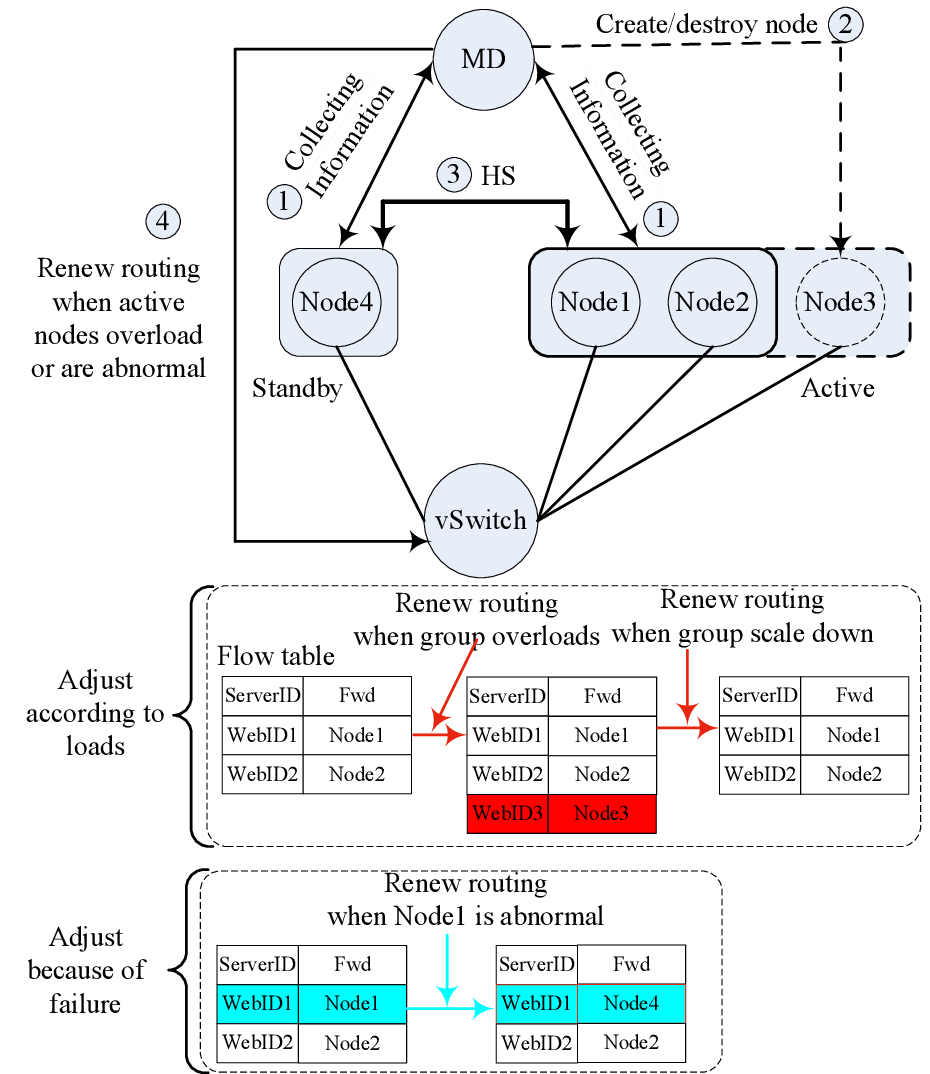}
\caption{The working mechanism of every group in SMG.}
\label{fig_sim5}
\end{figure}

When one group faces traffic overload, low-load and node failure, Fig 4 presents NetSecCC how to deal with them. In the face of overload and low load, MD collects and receives load and traffic information (e.g., session, cpu utilization) from active nodes in real time (1); According to the received information, MD makes such a determination that if the load of the active nodes is not balanced, MD renews routing in vSwitch to adjust the load between the active nodes (4). If all active nodes overload, MD creates a active node (2), and dynamically generates new routing to renew forwarding rules in vSwitch (4) to balance the load in active nodes. If active nodes appear low load, MD may destroy a active node (2), and renew forwarding rules in vSwitch (4) to improve resource utilization. To make load balancing contrate, we present from Fig 4 how to MD adjusts flow table to rebalance load. Initially, WebID1 and WebID2 traffic is forwarded to node1 and node2, respectively, to be inspected and filtered. If WebID3 traffic goes through node1 and node2, they overload. So MD changes this situation to create node3, and adds forwarding rules to route WebID3 traffic to node3. When WebID3 traffic ends, MD deletes rules forwarding traffic to node3, and destroys node3.

Since most middleboxs are stateful, a middelbox fails to result in loss of the established sessions in its memory. If a client accesses to the remote server again, the rebooted middlebox needs to re-establish the new session between the client and the server, resulting in a large time delay. Although the traditional HS is able to solve this problem by one-to-one switchover between active nodes and standby nodes, too many standby nodes seriously reduce resource utilization. Because middleboxs failure is small probability event, NetSecCC uses many-to-one mapping relationship between all active nodes and one standby node, that is, state information in all active nodes are synchronized to to the same standby node. When any active node fails, between the failed active node and the standby node achieve automatic switch-over, the standby node immediately takes on the role of the active node. At the same time, MD renew forwarding rules to route traffic to the switched standby node. The improved HS not only overcomes a long time delay to reboot the failed middlebox, but also improves system resource utilization. A specific example as shown in Fig 4 is that node1 suddenly fails, node4 immediately takes responsibilities of node1 by switch-over, and forwarding rules in vSwitch is renewed to route WebID1 to node4. This process is done automatically without human involvement. A more detailed process is shown in Section 3.2.

\section{Implement}
\label{sec:1}
The above design elaborates the principle of NetSecCC. In this section, we represent the implement of NetSecCC in detail. For the implementation of NetSecCC, we focus on the implementation of MD and security group. MD provides cloud users's services placed on cloud computing with their corresponding SIC to ensure their network security, and dynamically adjusts load balancing in security groups according to the load information. While security groups not only perform security inspection and filtering, but also improve quality of security service, including resource utilization, fault tolerance, high availability. We first demonstrate MD implement.

\subsection{MD Implement}
\label{sec:2}
MD as a SDN controller plays two important role in NetSecCC implementation. First, it controls traffic forwarding through their corresponding SIC of cloud users' services in service domains to ensure these services security. Second, it rebalances load due to traffic overloads or low load or node failure in each group. MD implements its two functions by forwarding rules in vSwitch, the implement process is shown in Fig 5. Resource manager sorts and analyzes these data from the inputs: state information from groups in SMG (e.g., CPU, session, etc), state information from vSwitch, groups topology and SIC (on-demand security service), and outputs the parameters considered as the inputs of RouteGen. RouteGen converts these parameters to forwarding rules using forwarding traffic through groups in SMG to be inspected and filtered. Until deployment of new services or security needs change in service domains, traffic overload or low load or node failure in groups, MD generates or renews rules in vSwitch.
\begin{figure}[!ht]
\centering
\includegraphics[width=2.5in]{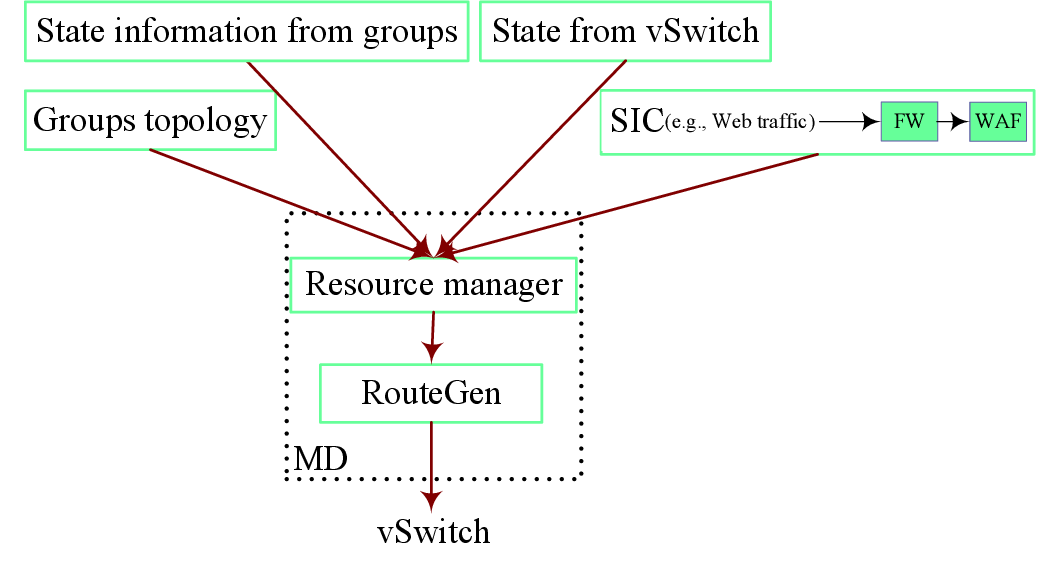}
\caption{MD implement}
\label{fig_sim4}
\end{figure}

\subsection{Security Group}
\label{sec:3}
Above we have introduced MD that is consider as a conductor for network security inspection and filtering of cloud computing, security groups are seen as the actual operator to put into effect specific security inspection. In this subsection, we elaborate on the implementation of security group in detail by means of focusing on load balancing and fault tolerance of each group in SMG as shown in Fig 6. We first dwell on their cooperation between MD, active nodes and vSwitch to implement load balancing between active nodes, Fig 6(a) present their work sequence and communication.
\begin{figure*}[bhtp]
\begin{minipage}{0.55\linewidth}
  \centerline{\includegraphics[width=3.2in]{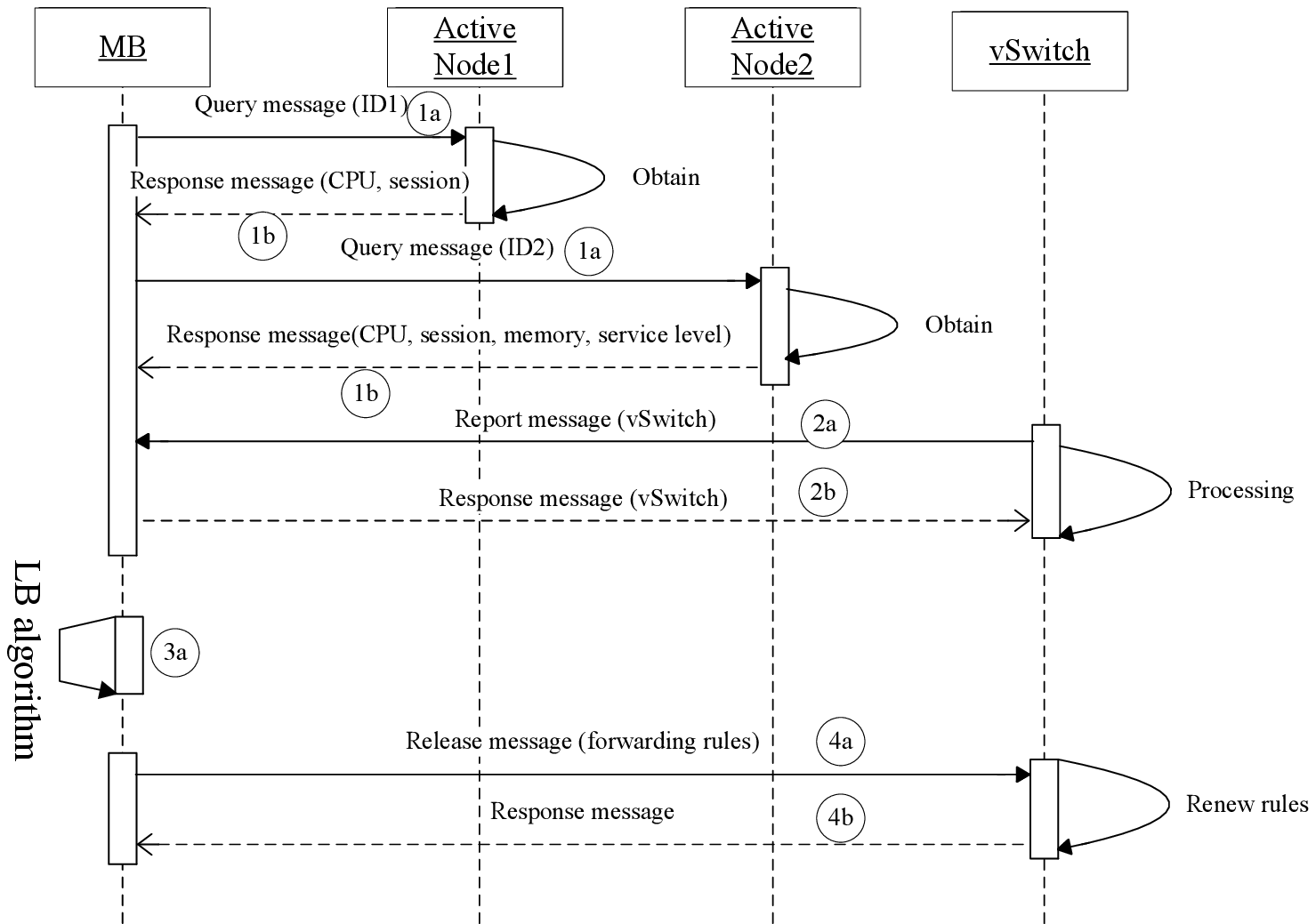}}
  \centerline{(a) Scaling out and in WAF due to traffic overload and low load.}
\end{minipage}
\hfill
\begin{minipage}{0.50\linewidth}
  \centerline{\includegraphics[width=3.1in]{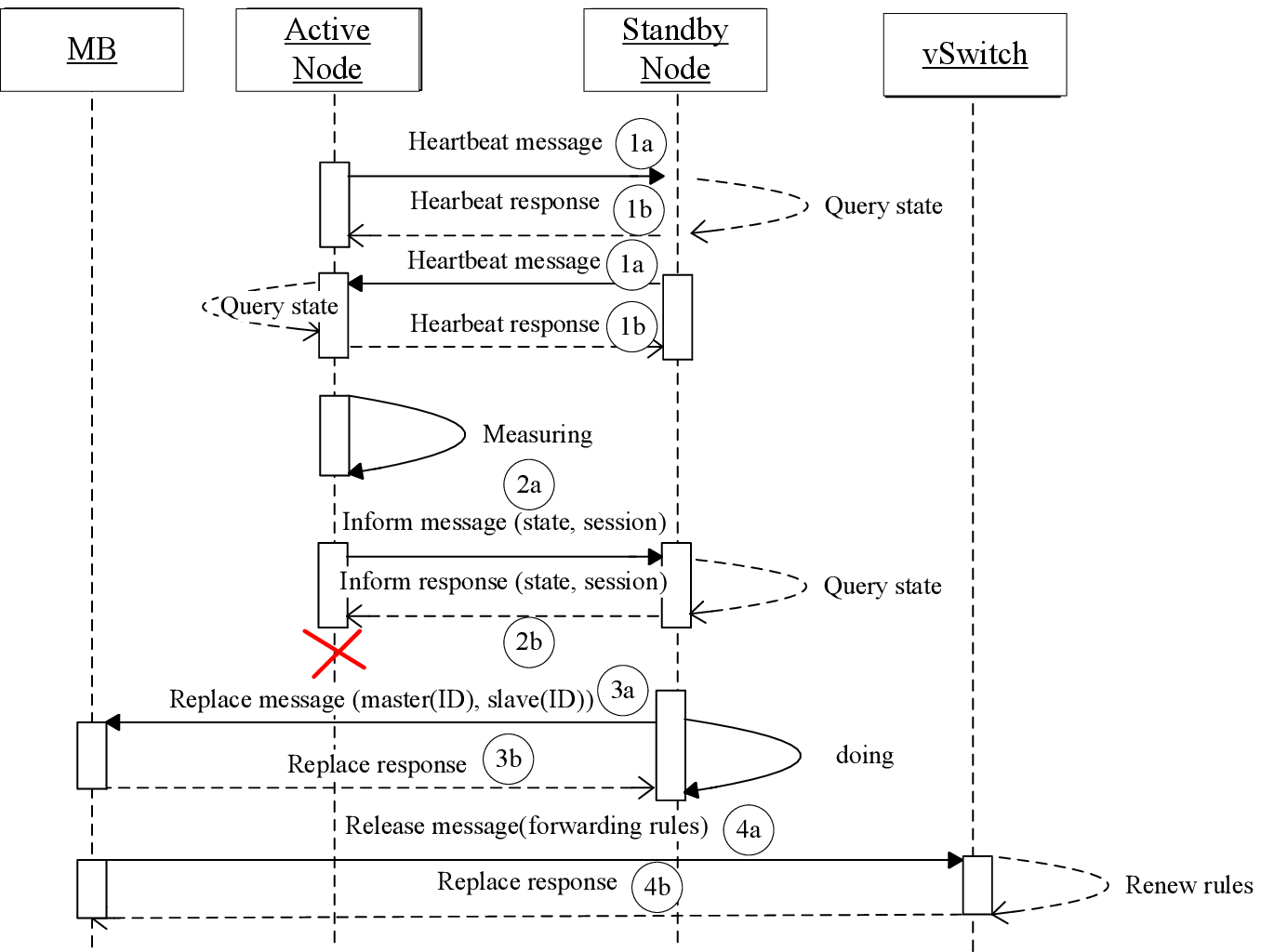}}
  \centerline{(b) Fault-tolerance processing.}
\end{minipage}
\caption{Load balancing and fault-tolerance processing in one group.}
\label{fig_sim7}
\end{figure*}

\begin{enumerate}
  \item MD$\xrightarrow [1a]{ReqMessage_{query}(ID_{Node})}$active Nodes: $ID_{Node}$ denotes the current active node identifier, query message queries all active nodes' state information.
  \item Active nodes$\xrightarrow [1b]{ResMessage_{query}(ID_{Node}, CPU, memory)}$MD: All current active nodes respond to their own state information including CPU utilization, memory usage, sessions, etc.
  \item vSwitch$\xrightarrow [2a]{ReqMessage_{report}(vSwitch)}$MD: vSwitch periodically reports the number of packets and flow size through active nodes to MD.
  \item MD$\xrightarrow [2b]{ResMessage_{report}(vSwitch)}$vSwitch: MD responds report message to vSwitch.
  \item LB algorithm: Alg (the number of packets, flow size, CPU, memory, session, etc) (3a). Note: load balancing algorithm is used current popular algorithm such as round robin, dynamic server act, dynamic ratio-APM, etc. The article concentrates on security architecture of cloud computing, load balancing algorithm is not too much expressed.
  \item MD$\xrightarrow [4a]{ReqMessage_{release}(vSwitch)}$vSwitch: MD issues forwarding rules to vSwitch according to LB algorithm.
  \item vSwitch$\xrightarrow [4b]{ResMessage_{release}}$MD: vSwitch responds to MD for release message.
\end{enumerate}

Next, we elaborate on work sequence and communication between active nodes and standby node to prepare for fault-tolerant on account of any active node failure as show in Fig 6(b).
\begin{enumerate}
  \item Active or standby node$\xrightarrow [1a]{ReqMessage_{heartbeat}(ID_{Node})}$standby or active node: They probe each other to determine whether the other is alive.
  \item Standby or active node$\xrightarrow [1b]{ResMessage_{heartbeat}(ID_{Node})}$active or standby node: They respond to each other probes.
  \item Active node$\xrightarrow [2a]{ReqMessage_{inform}(ID_{Node}, state, session)}$standby node: The renewed information (such as state, session) is backed up from active nodes to standby node in real time.
  \item Standby node$\xrightarrow [2b]{ReqMessage_{inform}(ID_{Node})}$active node: Standby node responds to active nodes for the renewed message.
\end{enumerate}

Finally, we explain Hot Standby switch-over process in accordance with an active node failure and present sequence and communication between MD and standby node to implement fault-tolerant as show in Fig 6(b), namely, any active node fails, the standby node immediately takes on its responsibilities.
\begin{enumerate}
  \item Standby node$\xrightarrow [3a]{ReqMessage_{replace}(ID_{active}, ID_{standby})}$MD: Standby node has probed that the active node has occurred an exception, then sends a replacement message to MD to switch over between the active node and the standby node.
  \item MD$\xrightarrow [3b]{ResMessage_{replace}}$Standby node: MD responds to the standby node for switch-over.
  \item MD$\xrightarrow [4a]{ReqMessage_{release}(vSwitch)}$vSwitch: MD renews forwarding rules to vSwitch after switch-over. The standby node becomes one active node to be responsible for inspecting and filtering the received traffic.
  \item vSwitch$\xrightarrow [4b]{ResMessage_{release}}$MD: vSwitch responds to MD for the renewed message.
\end{enumerate}

To summarize, MD guilds security groups to put into effect security inspection for traffic accessing to service domains, ensuring that traffic arriving at service domains is secure and trusted, while security groups are the specific implementer of security inspection. They complement each other to achieve security protection of cloud computing.

\section{Evaluation}
\label{sec:1}
In this section, We evaluate NetSecCC with the following goals:
\begin{itemize}
\item evaluate NetSecCC¡¯s ability to provide dynamic scalability to complex real world middleboxes, and measure the gain in resource utilization when scaling in a deployment using NetSecCC (\S 4.1).
\item evaluate the response time, especially fault-tolerant time and creation VM time when one or more of the active nodes fail (\S 4.2).
\item evaluate system overhead with NetSecCC compared to without security protection in cloud computing (\S 4.3).
\end{itemize}
\begin{table}[!ht]
\renewcommand{\arraystretch}{1.3}
\caption{The list of open source software about security softwares}
\label{table_5}
\centering
\begin{tabular}{r|l}
\hline
\hline
\bfseries Product Name & \bfseries Open Source Software \\
\hline
FW & IPFire \cite{FW} \\
WAF & ModSecurity \cite{WAF} \\
SSL/VPN & OpenSSL \cite{openssl} \\
AS & PacketFence \cite{MSG} \\
\hline
\hline
\end{tabular}
\end{table}
\textbf{Experimental environment} Cloud platform was conducted on a Dell Server with 8 core, 3.42 GHz Intel CPU, 16GB memory. The XEN hypervisor version is 3.4.2, the dom0 system is fedora 16 with kernel version 2.6.31. We used a 64bit fedora Linux with kernel version 2.6.27 as our guest OS, vSwitch bandwidth is 1 Gigabit Ethernet; NetSecCC uses open source security as softwares shown in TABLE I. Note that in our experiments, we use open source software about the network security rather than middleboxs from big security vendors, thus easily migrating these security software to cloud
computing.
\subsection{Scalability}
\label{sec:2}
\begin{figure}[!ht]
\begin{minipage}{1.00\linewidth}
  \centerline{\includegraphics[width=2.8in]{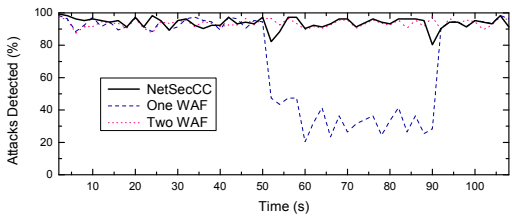}}
  \centerline{(a) Scaling out and in WAF due to traffic overload and low load.}
\end{minipage}
\\
\begin{minipage}{1.00\linewidth}
  \centerline{\includegraphics[width=2.8in]{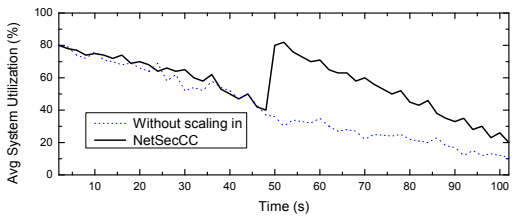}}
  \centerline{(b) System utilization when scaling in FW for low load.}
\end{minipage}
\caption{Scaling out and in security groups to test NetSecCC's scalability due to traffic overload and low load.}
\label{fig_sim7}
\end{figure}
Figure Fig 7(a) shows NetSecCC's ability to dynamically scale WAF out and in during a load burst. Our experiments environment is that Web server \cite{Apache-HTTP-Server} is installed in a separate VM in service domains, 30 clients in the form of a continuous sequence of POST requests access to Web server, the requests contain SQL injection and cross-site scripting attacks, and each client generates 80 requests/second. We inject a load burst 50 seconds into the experiment by introducing an additional 30 clients, the load burst lasts 40 seconds. We compare three scenarios: a single WAF instance that handles the entire load burst, a pair of WAF that share load (flows are assigned to each WAF in a round-robin fashion) and NetSecCC. NetSecCC scenario begins with a single WAF, when it overloads, NetSecCC creates a new WAF to split Web traffic.

As shown in Figure 7(a), until the load burst at t = 50s, all three scenarios have a 100\% detection rate. During the load burst, the performance of the single WAF reduces drastically because packets are dropped and attacks are missed. The two WAF do not experience any degradation as it has enough capacity and the load is well balanced between the two WAF. While NetSecCC creates a new WAF to split Web traffic according to the load burst at t=50s, this leads to this problem: packets are dropped and attacks are missed. However, the detection rate quickly rises because the two WAF have enough capacity for the load burst. After the load burst (t = 90s), NetSecCC detects a drop in load according to destroying one WAF. NetSecCC therefore enables WAF to handle the load burst without wasting resources by running two WAF throughout the entire experiment.

Figure 7(b) shows system utilization between NetSecCC and a pair of FW that share load (flows are assigned to each FWs in a round-robin fashion). Our experiments environment is that UDP server is installed on a separate VM in service domains, 100 UDP clients continuously send UDP packets to UDP server, each client evenly generates from 8M requests/second to 1M requests/second within 100 seconds in the descending way, NetSecCC initially has two FW to share UDP traffic. NetSecCC system utilization from 80\% to 50\% is the same as this pair of FW in 50 seconds before, however, NetSecCC utilization burst reaches 80\% at t=50s, the main reasons that NetSecCC is configured with a scale-in policy that triggers once one FW load falls below 50, one FW is destroyed, another FW is responsible for consolidating resources during low load and improves overall system utilization. After 50 seconds, NetSecCC's system utilization decreases from 80\% to 20\% with less traffic, while this pair of FW reduce to 10\%.
\subsection{Fault Tolerance}
\label{sec:3}
We consider one dynamic scenario in Fig 4. When one of the active nodes fails, we need to rebalance the load and we are interested in the time to recover the normal running network. For failure, Figure 8 shows a breakdown of the time it takes to detect time (collect and receive state information), generate new rules, and install them. Here balancing the load only takes 1 second by MD. However, we have to consider a special case, if the other active nodes have been in overload, rebalancing the load between the active nodes may further worsen the entire system. It is necessary to create a new active node to split traffic, which leads to a longer response time (3 seconds) as indicated CreateNode in Fig 8, this is unacceptable to cloud users. Switch-over of HS spends little time (1.2 seconds), which is far lower than CreateNode. So when one or more of the active nodes fail, MD divided into two cases: if the other active nodes overload, then MD launches HS, otherwise, activates rebalancing the load.
\begin{figure}[!ht]
\centering
\includegraphics[width=2.8in]{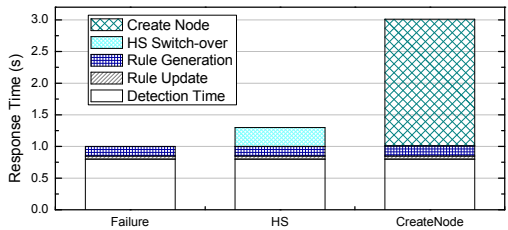}
\caption{Response time in the case of a middlebox failure}
\label{fig_sim4}
\end{figure}

\subsection{Performance Overhead}
\label{sec:4}
To evaluate NetSecCC's system performance overhead, we use throughput and latency that are important indicator of system performance as an evaluation criteria. Although this way without employing NetSecCC is higher efficiency than one with NetSecCC, if we do not make the protective measures to protect cloud computing security, this may lead to incalculable losses. So we have to protect network security of cloud computing to defend various attacks from the network. Even if we select NetSecCC to protect cloud computing security, we also have to consider whether its performance overhead can be accepted. We use IXIA \cite{IXIA} as a performance testing tool to evaluate NetSecCC's performance overhead, comparing both with and without network security in cloud computing. Next, we use two experiments to evaluate the performance impact with NetSecCC.

For Web page access as our first experiment, we use IXIA both as a customer and as a server to test with and without NetSecCC. The experimental results show in Fig 9, it is easy to see that NetSecCC has little impact on system performance for web page access, the average cost of its latency is 9.3\% (ranging from 6.4\% to 13.9\%) compared to without NetSecCC, the average cost of its throughput is 0.4\% (ranging from 0 to 3.7\%). The main reason for these overhead is that because SIC of web page access as shown Fig 3 is composed of FW group and WAF group, Web traffic must go through FW group and WAF group to be inspected and filtered, then arrives at Web server, while Web traffic directly accesses Web server without NetSecCC to avoid inspection with system overhead. So compared to without NetSecCC, latency becomes longer with NetSecCC, throughput is suffered from the impact of latency, but overall system performance with NetSecCC is within the acceptable.
\begin{figure}[bhtp]
\begin{minipage}{1.00\linewidth}
  \centerline{\includegraphics[width=2.6in]{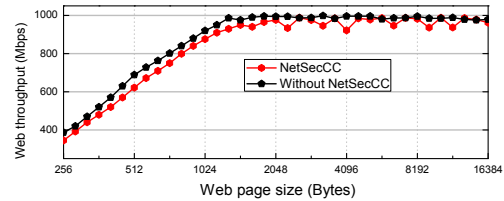}}
  \centerline{(a) Throughput comparison of web page access.}
\end{minipage}
\\
\begin{minipage}{1.00\linewidth}
  \centerline{\includegraphics[width=2.6in]{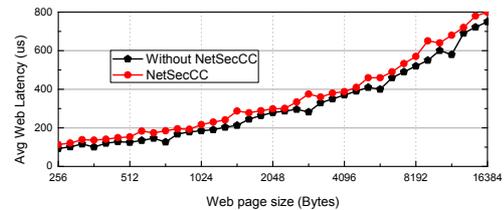}}
  \centerline{(b) Latency comparison of web page access.}
\end{minipage}
\caption{Performance comparison results between NetSecCC and without NetSecCC by Web page access.}
\label{fig_sim7}
\end{figure}

For Email access as our second experiment, we also use IXIA to test performance overhead with NetSecCC. The experimental results show in Fig 10, we can observe that even the encrypted emails with NetSecCC are only slightly affected. The impact of latency and throughput with NetSecCC is mainly due to emails must route through FW group, AS group and SSL/VPN group as shown in Fig 3, where the encrypted processing through SSL/VPN group consumes some time. Compared to without NetSecCC, specific data on the performance overhead with NetSecCC is shown below: the average cost of latency is 11.1\% (ranging from 9.2\% to 13.7\%), the average cost of throughput is 5\% (ranging from 0 to 11.1\%), such the performance overhead are perfectly acceptable relative to security services.
\begin{figure}[bhtp]
\begin{minipage}{1.00\linewidth}
  \centerline{\includegraphics[width=2.6in]{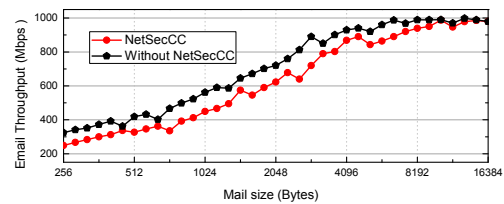}}
  \centerline{(a) Throughput comparison of mail access.}
\end{minipage}
\\
\begin{minipage}{1.00\linewidth}
  \centerline{\includegraphics[width=2.6in]{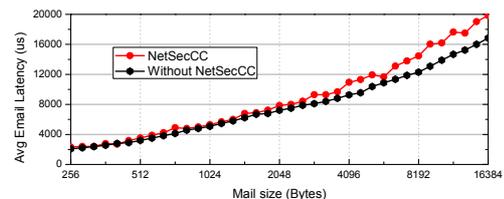}}
  \centerline{(b) Latency comparison of encrypted mail access.}
\end{minipage}
\caption{Performance comparison results between NetSecCC and without NetSecCC by mail access.}
\label{fig_sim7}
\end{figure}

In summary, by the comparison both with and without NetSecCC in cloud computing, NetSecCC is not only able to provide adequate network security protection for cloud computing, but also does not sacrifices the high price of system performance. The two experiments have further indicated that NetSecCC scheme can provide the efficient comprehensive network protection for cloud computing.

\section{Conclusion}
\label{sec:1}
Cloud users' services in cloud computing face network attacks from external and internal traffic. Both the traditional architecture and cloud providers and academia do not provide a novel flexible effective architecture to solve this problem. In this paper, NetSecCC is not only able to prevent from external and internal attacks to ensure cloud computing security, and provides flexible scalability and high effective fault-tolerant capability, but also experiments further show that NetSecCC's performance overhead is within the acceptable.

\begin{acknowledgements}
This work is partially support by the Open Research Project of the State Key Laboratory of Industrial Control Technology, Zhejiang University, China (No. ICT1407), JSPS KAKENHI Grant Number 25880002, 26730056 and JSPS A3 Foresight Program.
\end{acknowledgements}

\bibliography{reference}
\bibliographystyle{plain}

\end{document}